\documentclass[journal]{IJIT}
\usepackage{graphics,graphicx}

\begin{document}

\title{Pattern discovery for semi-structured web pages using bar-tree
representation}

\author{Z. Akbar, L.T. Handoko
        \thanks{All authors belong to the Group for Theoretical and
Computational Physics, Research Center for Physics, Indonesian Institute of
Sciences, Kompleks Puspiptek Serpong, Tangerang 15310, Indonesia, contact email
: gftk@mail.lipi.go.id.

Z. Akbar is also with the Group for Bioinformatics and Information
Mining, Department of Computer and Information Science, University of Konstanz,
Box D188, D-78457 Konstanz, Germany.}}

\maketitle
\thispagestyle{empty}

\begin{abstract}
Many websites with an underlying database containing structured data provide the
richest and most dense source of information relevant for topical data
integration. The real data integration requires sustainable and reliable pattern
discovery to enable accurate content retrieval and to  recognize pattern changes
from time to time; yet, extracting the structured data from web documents is
still lacking from its accuracy. This paper proposes the bar-tree representation
to describe the whole pattern of web pages in an efficient way based on the
reverse algorithm.  While previous algorithms always trace the pattern and
extract the region of interest from  \textit{top root}, the reverse algorithm
recognizes the pattern from the region of interest to both top and bottom roots 
simultaneously.  The attributes are then extracted and labeled reversely from
the region of interest of targeted contents. Since using conventional
representations for the algorithm should require  more computational power, the
bar-tree method is developed to represent the generated patterns using bar
graphs characterized by the depths and widths from the document roots. We show
that this representation is suitable for extracting the data from the
semi-structured web sources, and for detecting the template changes of targeted
pages. The experimental results show perfect recognition rate for template
changes in several web targets.
\end{abstract}


\IEEEpeerreviewmaketitle

\section{Introduction}
\label{intro}

Text mining, especially from the web sources is getting important during the
last decade. This is triggered by the exponentially growing number of websites
with various types of information on the net. Most of them are providing the
information generated from the structured data in an underlying database through
certain predefined templates or layouts \cite{cnhsu}. Following the great number
of web pages in this kind which are already available on the net, these
semi-structured web sources contain rich and unlimited valuable data for a
variety of purposes. Extracting those data and then rebuilding them into a
structured database are a challenge to realize an automatic data mining from web
sources. 

Several methods for these purposes have been proposed previously in the
literature. Some of them can be classified as the so-called wrappers
\cite{w1,w2,w3,w4,w5} which have been briefly surveyed in \cite{w6}. The wrapper
technique allows an automatic data extraction through predefined wrapper created
for each target data source. The wrappers then accepts a query against the data
source and returns a set of structured results to the calling application. 

On the other hand, there are several automatic methods without a manual initial
learning process. For example, some methods are generating the template
automatically from first multiple pages before extracting the rest of data based
on the template \cite{t1,t2,t3}. A more comprehensive method without requiring
multiple pages has also been proposed using a page creation model which captures
the main characteristics of semi-structured web pages to derive the set of
properties \cite{ti1}. Though, the last method is more intended to extract the
lists of data records from a single web page with sibling subtrees.

Since 2008 our group has worked on developing an online infrastructure with a
major purpose of integrating the information related to science and technology
across Indonesia, that is the Indonesian Scientific Index -- ISI \cite{isi}.
However, in contrast with conventional approach where the data are collected
through official connection under certain regulation, ISI integrates the data
indirectly by harvesting certain web contents of the official websites of 
targeted institutions. Initially the method was motivated by the failures of
some  \textit{conventional} methods of data integration which always requires
certain standard at any level and leads to additional works in the participating
institutions. On the other hand, as a part of public responsibility all academic
institutions have developed and launched various public information through
their own websites. Therefore the idea of indirect data integration is welcomed
by all participating institutions, since it is not like a dictatorship, more
acceptable, much cheaper and more efficient for all parties than the
conventional one which requires a kind of standardization among the information
islands belong to  separated institutions \cite{coexist,isipaper}. The main
problem is yet improving the accuracy of data retrieval and restructuring them
into desired fields for further content analysis.

\begin{figure*}[t!]
 \centering
 \includegraphics[width=\textwidth]{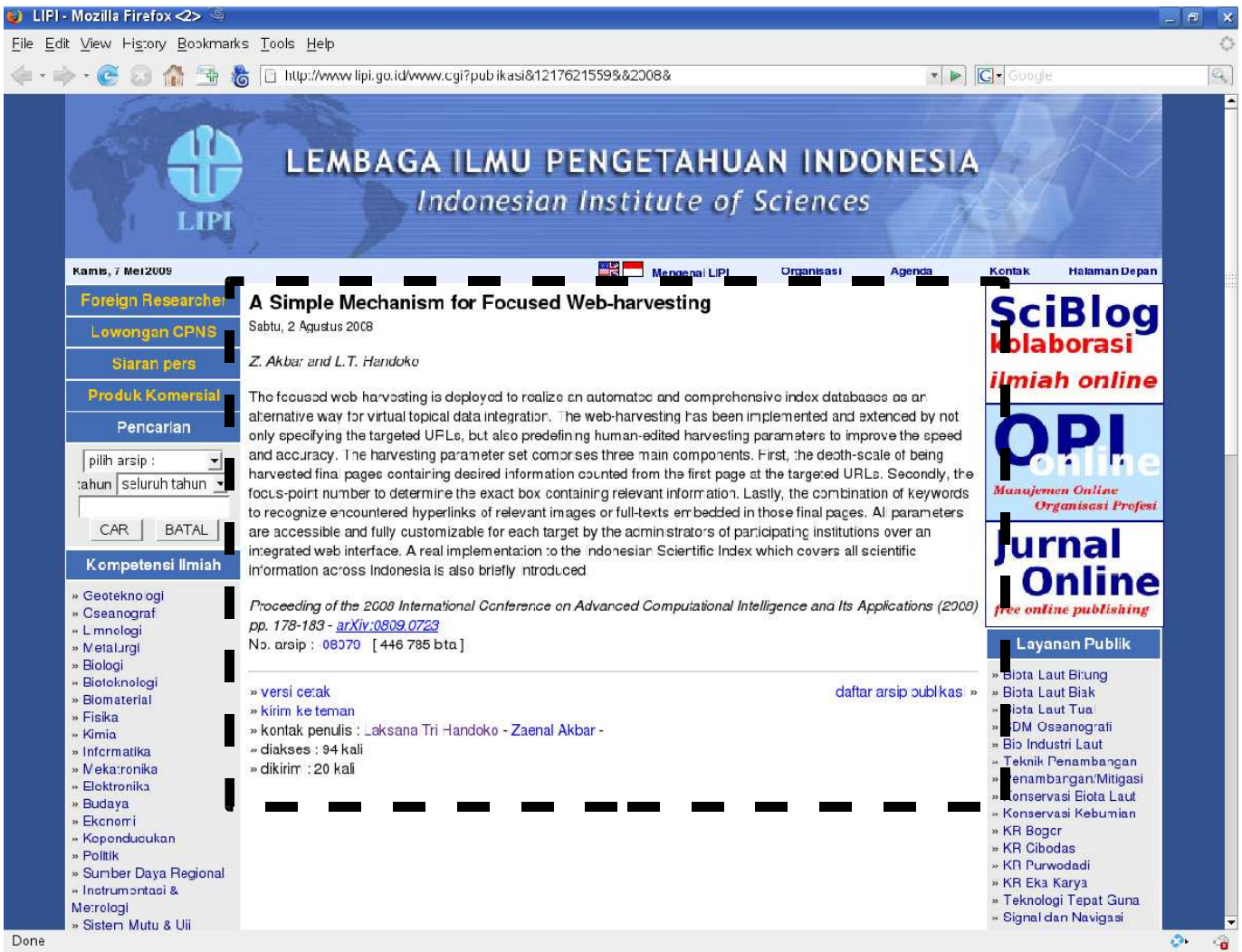}
 \caption{The example of the desired RoI from the displayed content of a
scientific publication page on the web, shown by the areas inside the dashed
rectangular.}
 \label{fig:source}
\end{figure*}

\begin{figure*}[t!]
 \centering
 \includegraphics[width=\textwidth]{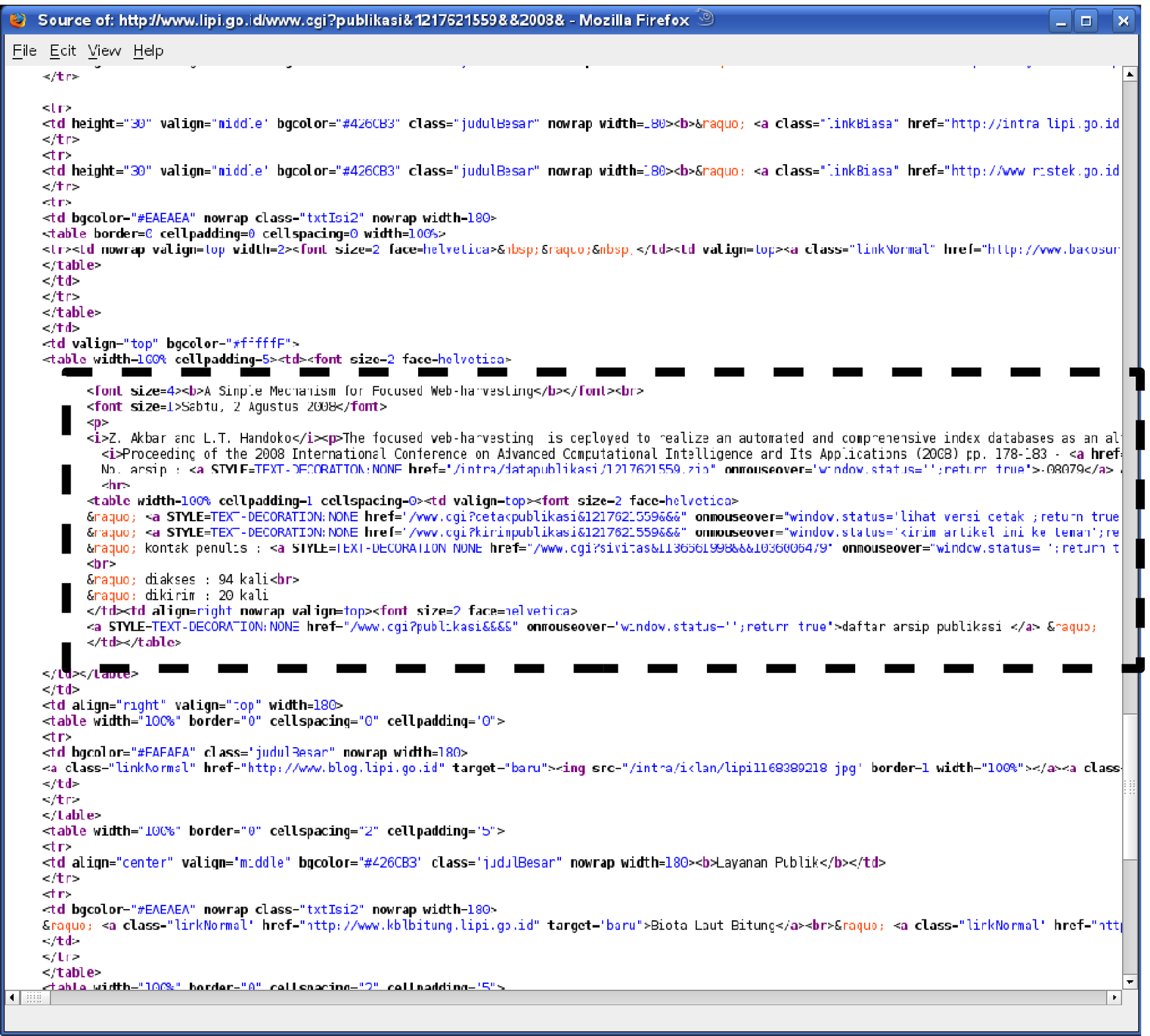}
 \caption{The example of the desired RoI from the web source of a scientific
publication page on the web in Fig. \ref{fig:source}, shown by the areas inside
the dashed rectangular.}
 \label{fig:source2}
\end{figure*}

The architecture of ISI is inspired and the combination of focused web-crawling
and regular web-harvesting. The focused web-crawling does not indiscriminately
crawl the web pages like general purpose search engines, but attempts to
download pages that are similar to each other \cite{chakrabarti}. Here, we
follow the same line to harvest certain types of web pages with specific
contents. Throughout the paper, let us call this method as the focused
web-harvesting \cite{isipaper}. Nonetheless, as an official data integrator it
must not allow any errors in the data collection. According to the data
integration purposes and its requirement of high accuracy, the first version of
focused web-harvesting technique at ISI adopts human intervention in the initial
setup by providing the targeted URL of list of data, for instance the
publication list, and defining the template of the final page contains the
relevant information by labeling the attributes. In the case of detail
information of a publication page as shown in Figs. \ref{fig:source} and
\ref{fig:source2}, the relevant information and labeling are  ranging from
title, authors, abstract to the full-text if available. Comparing with the
general purpose crawling \cite{crawl}, or even focused web-crawling
\cite{webcrawl}, obviously ISI could retrieve the information more accurately
due to its targeted contents and sources. The approach has also been realized
and integrated in a user friendly web-based interface to enable the
administrator to set up the initial parameters for each targeted sources. The
toolkit has been released as an open source under GNU Public License at
SourceForge.net \cite{openisi}.

Unfortunately, in spite of its high accuracy and unnecessary machine-learning
like system, the method is suffered from tedious labor time at its initial setup
to determine the region of interest (RoI), the tokens and to label the
attributes. Moreover, it is lacking of the ability to detect effectively later
changes of the targeted page templates. Any later changes of the targeted
templates will again require human intervention to manually revise the
parameters accordingly. Especially, because the labels and tokens are
represented as DOM trees which are sensitive to later changes of targeted web
templates \cite{t3,dom}. On the other hand, we have also found that the
previously known methods like wrapper induction \cite{w2}, or the information
extraction based on multi pattern discovery techniques \cite{chang}, are not
suitable for our purpose since all of them by definition contain significant
statistical errors which could burdens the initial purpose of official data
integration.

In our recent work  \cite{reverse}, in order to overcome the above-mentioned
problems a new method has been introduced to extract the RoI at the final
targeted web pages and its tokens reversely from the RoI to both top and bottom
roots, and further to label each attribute as usual. The technique is in
contrast with the existing mechanisms which always start from the top root of
targeted web pages or its RoI. It has been argued that the so-called reverse
algorithm is more efficient and accurate to extract the tokens, and on the other
hand to detect later template changes without any ambiguities. We should also
remark that the algorithm is applicable for any existing methods for data
extraction, especially the ones which require initial setup by human
intervention to define and label the attributes. This is actually similar to the
previous method \cite{chang}, but instead of using the PAT tree \cite{pat3} we
use the DOM tree like mechanism \cite{dom}. 

In this paper, we further present an alternative representation for pattern
discovery which is suitable for reverse algorithm mentioned above. The pattern
is described using simple bar graphs characterized by its widths, depths and the
partial and total squares.

The paper is organized as follows. First, after this brief introduction we
review shortly the reverse mechanism approach in Sec. \ref{sec:tree}. In the
subsequent section, Sec. \ref{sec:bar}, we present the so-called bar-tree
representation and its formalism. Finally we provide the experimental results of
deploying the method to detect the template changes before summarizing the paper
and discussing some future issues and further development.

\section{The reverse algorithm}
\label{sec:tree}

No matter the method used to extract and to label the tokens from a web template
or layout, correct initial setup is crucial for further data extraction from web
sources. As mentioned before, this point plays an important role for indirect
data integration which has no tolerance to any errors. This makes some methods
based on the machine learning system are useless. 

From now, please note that we are not going to discuss the algorithms to mine
the labeled data since the tasks after labeling can be done further using any
existing methods, nor to find the relevant pages of data list which has been
discussed in our and many others' previous works \cite{isipaper}. The reverse
mechanism can be outlined recursively as follows :
\begin{enumerate}
\item Determine the URL address of the final web page with desired information
like Fig. \ref{fig:source}.
\item Provide the whole sentences of the RoI by copying and pasting the 'desired
text' displayed on screen, not its source.
\item Provide the whole sentence(s) of each sub-RoI and assign the attributes
for each of them.
\item Crawl the source of the final web page at 1.
\item Parse and clean the text-format HTML tags like \verb|<b>|, \verb|<i>|,
etc.
\item Take the upper part of source from the top till the last one before the
first sentence of RoI. Parse and clean all texts inside except the layout-format
HTML tags, like \verb|<tr>|, \verb|<span>|, etc, Do the same thing for the lower
part that is from the end of last sentence of RoI till the bottom.
\item Count the number of 'open-tag' ($n_\mathrm{ot}$) and 'closed-tag'
($n_\mathrm{ct}$) from the deepest part in term of desired content, that is the
nearest tags from the RoI.
\end{enumerate}
We should stress here that there is no need for the administrators to provide
the web page sources at all. Open-tag here means the tags which have no pair in
upper or lower part, while the closed-tag denotes the pairing tags within upper
or lower part. Of course, our interest is only in the open-tag which should
describe the whole structure of web template.

\begin{figure*}[t]
 \centering
 \includegraphics[width=\textwidth]{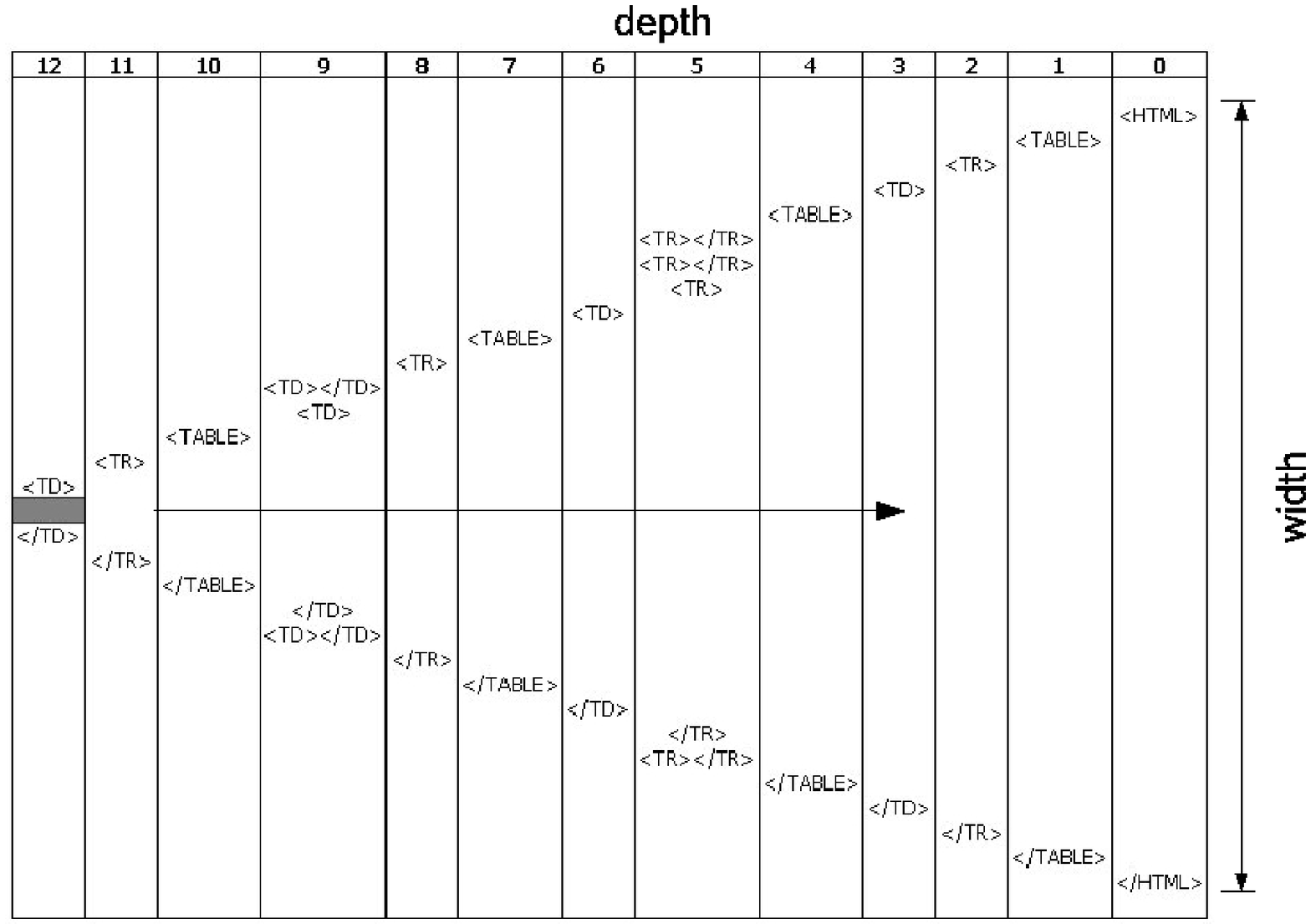}
 \caption{The tree for the example of RoI given in Fig. \ref{fig:source}. The
dashed box denotes the RoI.}
 \label{fig:tree}
\end{figure*}

Following the above procedures, we can obtain a kind of DOM tree as shown in
Fig. \ref{fig:tree}. We can further count the number of trees according to the
number of open-tags. Please remind that the counting is done horizontally, from
left to right shown by the arrow in the figure. The number of trees in upper and
lower parts are determined by, 
\begin{equation}
 \Sigma \equiv n_\mathrm{ot} - n_\mathrm{ct} \, .
\end{equation}

Concerning all possibilities on the number of trees in upper and lower parts,
therefore we can generally categorize the web structures through the
discrepancies between both numbers as,
\begin{eqnarray}
 \Delta & & \equiv \Sigma_\mathrm{upper} - \Sigma_\mathrm{lower} 
\nonumber \\
 & & \left\{ 
\begin{array}{lcl}
 = 0 	& : & \mathrm{fully \; symmetry} \\
 < 0 	& : & \mathrm{lower \; asymmetry} \\
 > 0 	& : & \mathrm{upper \; asymmetry} \\
\end{array}
\right.
	\label{eq:delta}
\end{eqnarray}
Fig. \ref{fig:tree} provides an example of tree in the case of Fig.
\ref{fig:source} which is accidently asymmetry. That means the number of trees
in the upper and lower parts are not the same, $\Sigma_\mathrm{upper} \neq
\Sigma_\mathrm{lower}$. Again, we can use one of the existing methods to
calculate the number of trees like  the PAT tree algorithm \cite{pat3} and so
forth. 

Through the discussion above, it is clear that the present method has several
advantages :
\begin{itemize}
 \item We can separate independently the structure and the rules to obtain the
RoI and the structure inside. 
\item We can find out the template changes and its relevance with the desired
RoI, since we can compare and see the pairing tokens between the upper and lower
parts.
\item There is no need for further human intervention as long as the page
containing the initial RoI still exist. The system uses the same RoI as keyword
to perform regular check to detect template changes before recrawl the same
target. Only if the content is removed by the owner, the system will defer the
recrawling job at the target and send a  warning to the administrator to choose
another content as new keyword.
\end{itemize}
We discuss these points in more detail through the real implementation at ISI in
the subsequent Sec. \ref{sec:implementation}.

Further issue is then how to represent the method, not only visually, but also
mathematically to enable more quantitative analysis in real implementation.

\section{The bar-tree representation}
\label{sec:bar}

\begin{figure*}[t]
 \centering
 \includegraphics[width=\textwidth]{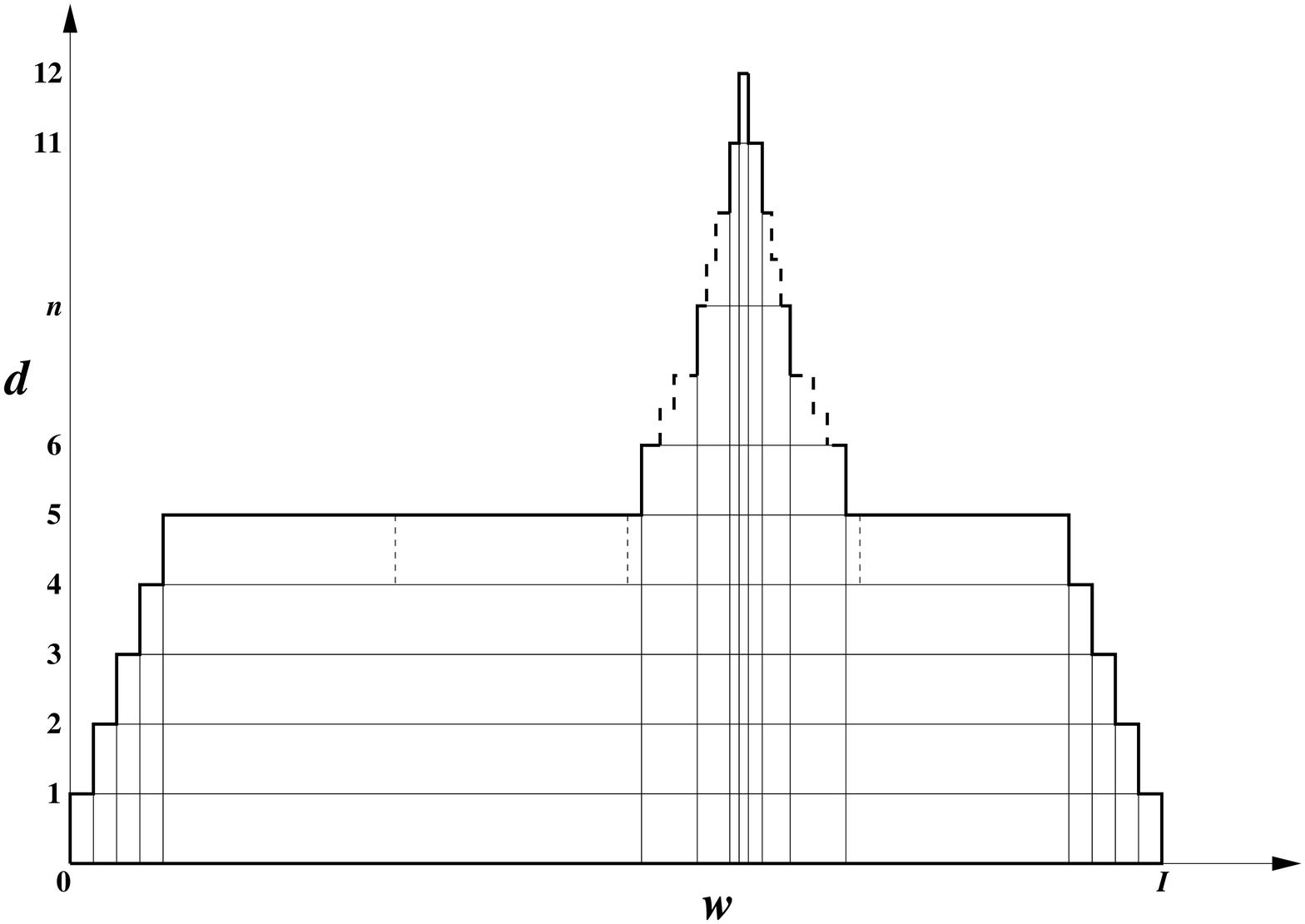}
 \caption{The bar-tree representation with $w$ and $d$ denote the width and
depth of each tree as depicted in Fig. \ref{fig:tree} for $r = 10\% \times I$.}
 \label{fig:box}
\end{figure*}

Here we introduce the way to represent the reverse algorithm method in form of
bar diagrams. The bar is characterized by its width ($w$) and height. The height
is determined by the depth ($d$) of each column of attributes from the root
document. On the other hand, the width is given by the number of parallel
attributes ($P_d$) in certain depth weighted by a ratio ($r$) according to the
depth. The definition is illustrated in Figs. \ref{fig:tree} and \ref{fig:box}.

Considering the mentioned-above definition, the width of each bar diagram can be
written as follows,
\begin{equation}
 	w_d = \left\{
	\begin{array}{lcl}
	I & \mathrm{for} & d = 0 \\
	\frac{\displaystyle I - (d - 1) \, r}{\displaystyle P_{d-1}} \, w_{d-1}
& \mathrm{for} & d > 0 \\
	\end{array}
    	\right. \, ,
	\label{eq:wd}
\end{equation}
where $I$ is the given initial width and $r \leq I/{d_\mathrm{max}}$ is the
appropriate ratio to decrease the width following the depth of a bar. The
parallel attributes $P_d$ is nothing else than the number of attributes at the
same $n-$th depth. For instance,  the 5th column in Fig. \ref{fig:tree} has $P_5
= 4$, while at the 9th column $P_9 = 3$ and so forth.

\begin{figure*}[t]
 \centering
 \includegraphics[width=\textwidth]{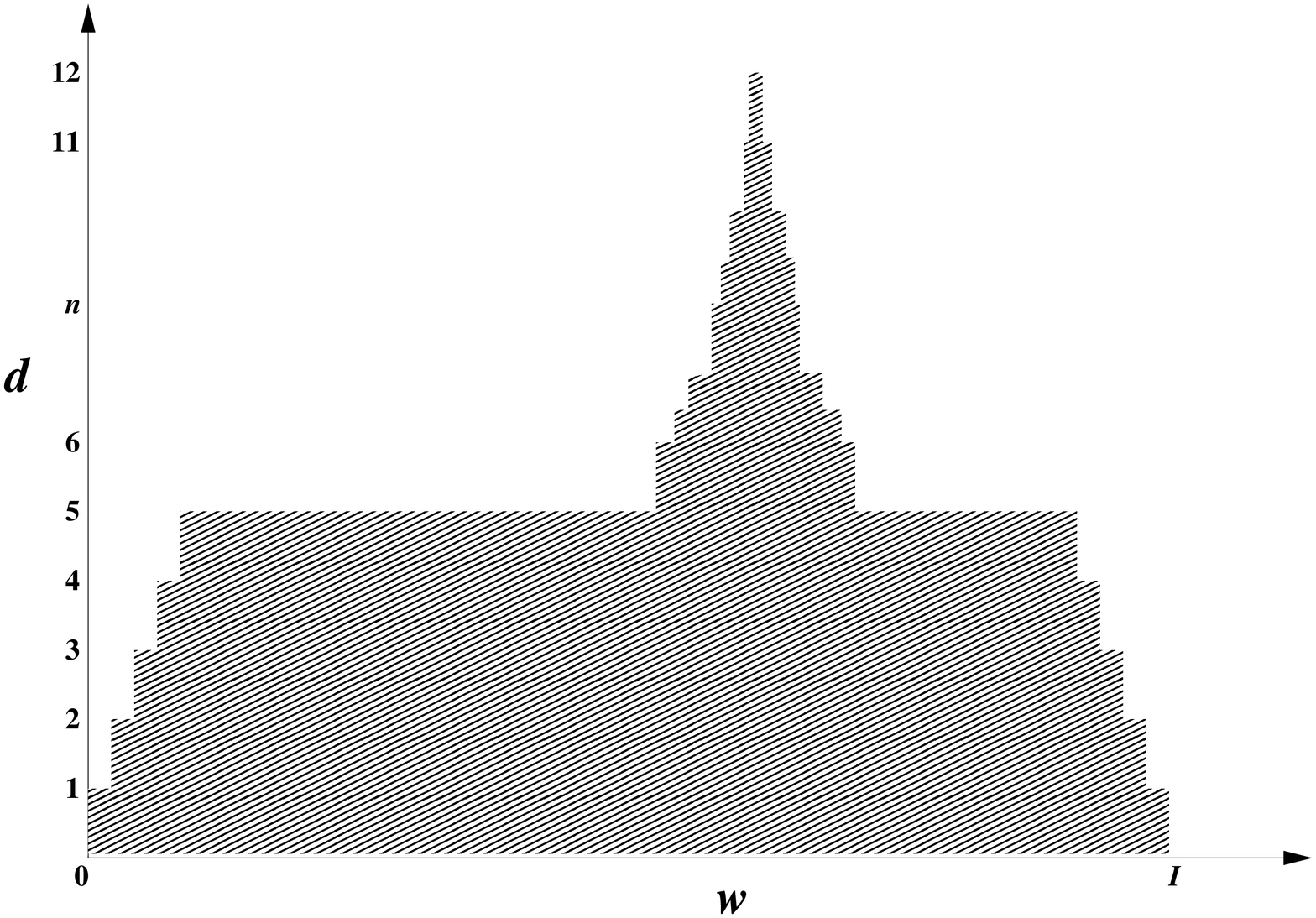}
 \caption{The total pattern of bar-tree representation in the case of Figs.
\ref{fig:tree} and \ref{fig:box}.}
 \label{fig:box2}
\end{figure*}

According to the definition, the bar-tree representation in the case of Fig.
\ref{fig:tree} can be further depicted as Fig. \ref{fig:box}. One should remark
a rule that for $P_d > 1$, the ($d+1$)th bar should be drawn inside the
appropriate order of bar in that  depth. In the case of $d=5$ in Fig.
\ref{fig:box}, since $P_5 = 4$ and the interested bar is the 3rd one, the
$d=6$'s bar is put inside the 3rd of $d=5$'s bar accordingly.

Quantitatively, the pattern of bar-tree representation can be characterized by
its partial and total squares. The square of each individual bar is given by,
\begin{equation}
 	A_d = \left\{
	\begin{array}{lcl}
	0	& \mathrm{for} & d = 0 \\
	d \times \frac{\displaystyle I - (d - 1) \, r}{\displaystyle P_{d-1}} \,
w_{d-1} & \mathrm{for} & d > 0 \\
	\end{array}
	\right. \, .
	\label{eq:ad}
\end{equation}

The ``nett-square``, that means the square of bar which is not overlapped with
its lower bars, for each bar is given by,
\begin{eqnarray}
	A_d^\mathrm{nett} & = & d \times \frac{I - (d - 1) \, r}{P_{d-1}} \,
w_{d-1} 
	- (d - 1) \times \frac{I - (d - 1) \, r}{P_{d-1}} \, w_{d-1}
	\nonumber \\
	& = & \frac{I - (d - 1) \, r}{P_{d-1}} \, w_{d-1} \; .
\end{eqnarray}
Using Eqs. (\ref{eq:wd}) and (\ref{eq:ad}), it subsequently leads to, 
\begin{equation}
	A_d^\mathrm{nett} = \left[ \prod_{n=0}^{d-1} \frac{I - (d - 1 - n) \,
r}{P_{d-1-n}} \right] \, w_0 \; .
	\label{eq:adr}
\end{equation}
Finally, the total square of pattern like Fig. \ref{fig:box2} becomes,
\begin{eqnarray}
	A^\mathrm{total} & = & \sum_{d=1}^{d_\mathrm{max}} \, A_d^\mathrm{nett} 
	\nonumber \\
 	& = & \sum_{d=1}^{d_\mathrm{max}} \left[ \prod_{n=0}^{d-1} \frac{I - (d
- 1 - n) \, r}{P_{d-1-n}} \right] \, w_0 \; .
	\label{eq:adt}
\end{eqnarray}
Since the initial width is arbitrary and determines only the normalization
factor for the square, one can simply put $w_0 = I = 1$ in real implementation. 

Now we have all formulae at hand and are ready to implement it to harvest the
information from web sources.

\section{Implementation}
\label{sec:implementation}

Our approach to web page information extraction has been experimentally
implemented into the system of ISI. Following the wrapper induction programs
which are usually supplemented by a user friendly GUI, we have also developed a
web based interface to perform the initial setup \cite{reverse}.

ISI can now efficiently detect the changes of targeted web templates by
comparing the old and new numbers of variable sets defined above as
$d_\mathrm{max}$, $A^\mathrm{total}$ and $A_d$. It is done by executing the
check procedure each time prior to the new crawling works to the same targeted
web pages. Once all variables and keys have been stored in the system, it can be
used not only for the initial setup but also for rechecking the templates from
the same web sources in a regular basis without human intervention.

In most cases, the discrepancies between the old and new numbers of variable set
$\{ d_\mathrm{max}, A^\mathrm{total} \}$ are already enough to detect easily the
template changes time by time. However, further discrepancies and locating its
details should be examined at each level of depth using $P_d$ and $A_d$. The
important point is, once the template changes have been detected, the system can
be designed to automatically replace the old version of template with the new
one without human intervention. However, if the content (of initial RoI) is
missing, because in most cases removed by the owner, the system can be designed
to defer the recrawling job at the target, and to send a kind of warning to the
administrator. Then, in this case the administrator should choose another
content as new keyword to locate the desired RoI.

We have done an experimental running on a Linux PC (Core 2 Duo 2.6 GHz
processor, memory 2 GB) to perform initial extraction and to detect later
template changes using 10,000 data from 20 different sources available at ISI.
The data belong to five categories with different maximum depths, \textit{i.e.}
$d_\mathrm{max} = (5,10,15,20,25)$. 

\begin{figure*}[t]
 \centering
 \includegraphics[width=\textwidth]{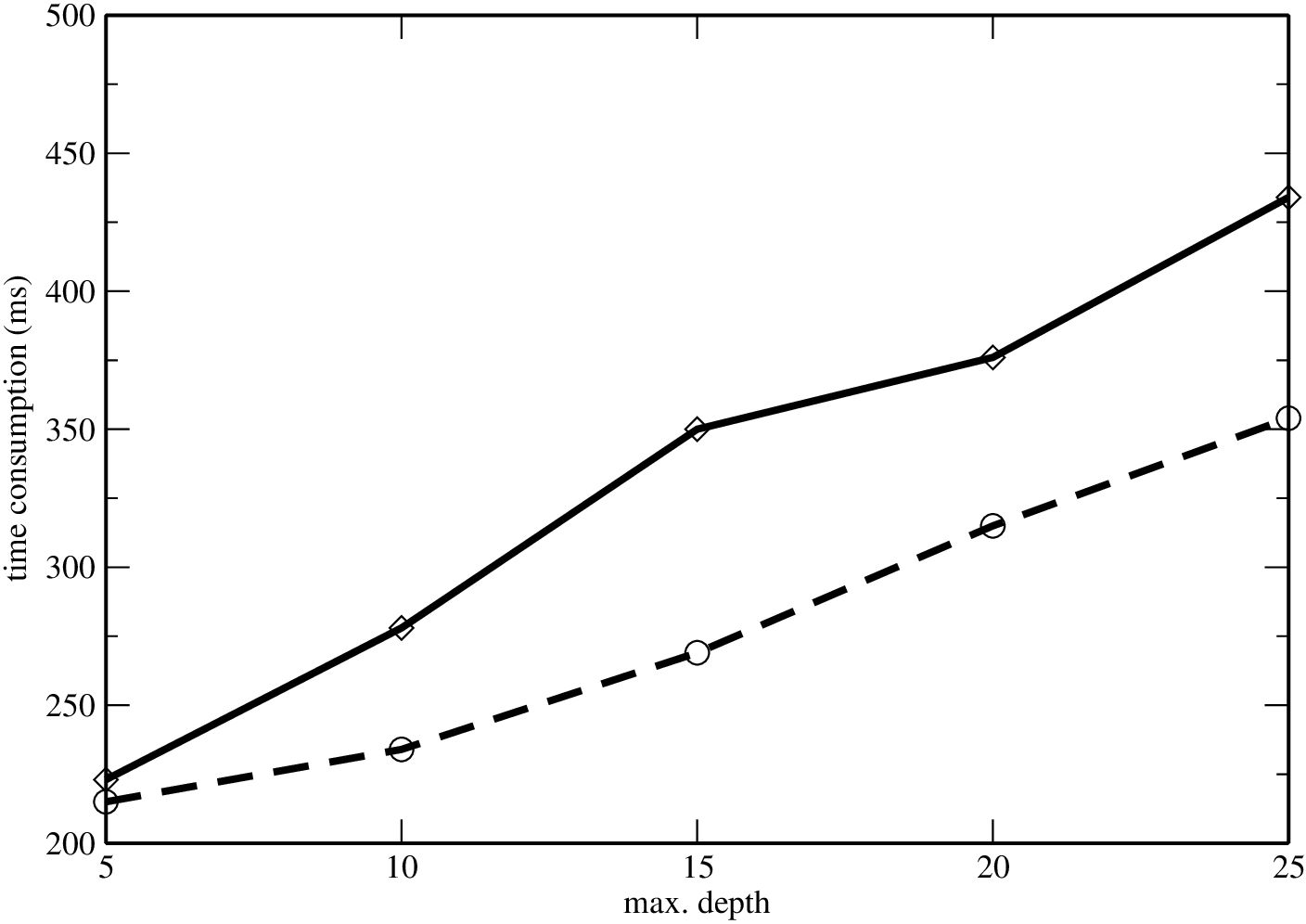} 
 \caption{The measurement result of time consumption for detecting the template
changes with various number of $d_\mathrm{max}$. The solid and dashed lines
denote the performance with variable sets : $\{ d_\mathrm{max},
A^\mathrm{total}, P_d, A_d \}$ and $\{ d_\mathrm{max}, A^\mathrm{total} \}$.}
 \label{fig:result}
\end{figure*}

\begin{figure*}[t]
 \centering
 \includegraphics[width=\textwidth]{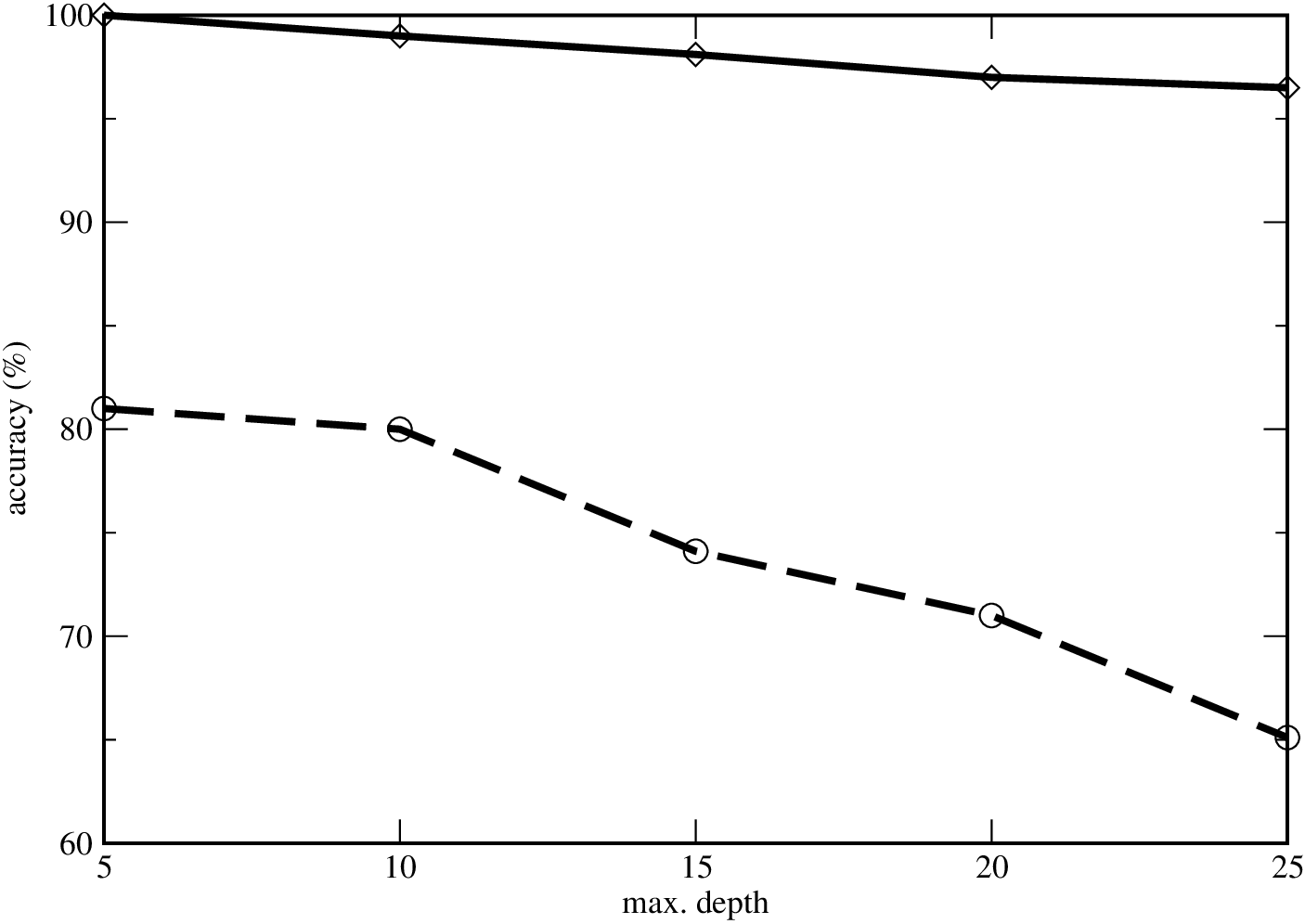}
 \caption{The measurement result of accuracy for detecting the template changes
with various number of $d_\mathrm{max}$. The solid and dashed lines denote the
performance with variable sets : $\{ d_\mathrm{max}, A^\mathrm{total}, P_d, A_d
\}$ and $\{ d_\mathrm{max}, A^\mathrm{total} \}$.}
 \label{fig:result2}
\end{figure*}

The results for two cases representing the calculation using different variable 
sets, $\{ d_\mathrm{max}, A^\mathrm{total} \}$ and $\{ d_\mathrm{max},
A^\mathrm{total}, P_d, A_d \}$, are shown in Figs. \ref{fig:result} and
\ref{fig:result2}. The time consumption in Fig.  \ref{fig:result} means the
running time in millisecond required for the whole processes from  parsing the
HTML till calculating the whole variables defined above. While the accuracy rate
in Fig. \ref{fig:result2} represents the successful rate to detect the template
changes. The template changes were done randomly but intensively among the data
using certain script. We should note that the template changes are completely
same for both variable sets. 

From the figures, we can deduce several points :
\begin{itemize}
\item As the number of $d_\mathrm{max}$ is greater, it requires more time to
calculate all variables and also decreases slightly the accuracy. The reason is
obvious, since the trees with deeper structure have more probabilities and
complexities of template changes.
\item Using more complete variable set $\{ d_\mathrm{max}, A^\mathrm{total},
P_d, A_d \}$  would improve significantly the accuracy to detect the template
changes than the simpler one $\{ d_\mathrm{max}, A^\mathrm{total} \}$ without
significant increasing in time.\\
Because the variables $A^\mathrm{total}$ could be accidently the same if the
template changes occurred at the same depth $d$. For instance, if only the
sequence in a depth $d$ is different, then $P_d$ should remain unaltered.
\item Although the variable set $\{ d_\mathrm{max}, A^\mathrm{total}, P_d, A_d
\}$ is enough in most applications, yet there is no guarantee to correctly
detect the location of template changes. Fortunately, it is indeed not necessary
in the reverse algorithm. Because once the template changes were detected, the
new pattern is re-extracted from the RoI to replace the old one, no matter where
the changes happened.
\end{itemize}

If one requires, for certain needs, more accurate detection power, that is 100\%
in our experimental running, we recommend to perform further check using the
variable $\Delta$ in Eq. (\ref{eq:delta}). The discrepancies between the old and
new numbers of $\Delta$ would unambiguously detect the template changes and its
exact location. It can be summarized as below,
\begin{enumerate}
 \item No change at all :\\
	$\Delta^\mathrm{new} = \Delta^\mathrm{old}$; 
	$\Sigma_\mathrm{upper}^\mathrm{new} =
\Sigma_\mathrm{upper}^\mathrm{old}$; 
	$\Sigma_\mathrm{lower}^\mathrm{new} =
\Sigma_\mathrm{lower}^\mathrm{old}$
\item Simultaneous changes with same size in both upper and lower trees : \\
	$\Delta^\mathrm{new} = \Delta^\mathrm{old}$; 
	$\Sigma_\mathrm{upper}^\mathrm{new} \neq
\Sigma_\mathrm{upper}^\mathrm{old}$; 
	$\Sigma_\mathrm{lower}^\mathrm{new} \neq
\Sigma_\mathrm{lower}^\mathrm{old}$ 
\item Only either upper or  lower tree has changed :\\
	$\Delta^\mathrm{new} \neq \Delta^\mathrm{old}$; 
	$\Sigma_\mathrm{upper}^\mathrm{new} \neq
\Sigma_\mathrm{upper}^\mathrm{old}$; 
	$\Sigma_\mathrm{lower}^\mathrm{new} =
\Sigma_\mathrm{lower}^\mathrm{old}$ \\
	or :  \\
	$\Delta^\mathrm{new} \neq \Delta^\mathrm{old}$; 
	$\Sigma_\mathrm{upper}^\mathrm{new} =
\Sigma_\mathrm{upper}^\mathrm{old}$; 
	$\Sigma_\mathrm{lower}^\mathrm{new} \neq
\Sigma_\mathrm{lower}^\mathrm{old}$ \\
\item Both upper and lower trees have changed differently :\\
	$\Delta^\mathrm{new} \neq \Delta^\mathrm{old}$; 
	$\Sigma_\mathrm{upper}^\mathrm{new} \neq
\Sigma_\mathrm{upper}^\mathrm{old}$; 
	$\Sigma_\mathrm{lower}^\mathrm{new} \neq
\Sigma_\mathrm{lower}^\mathrm{old}$ \\
\end{enumerate}
Apparently, in the case 1 no need to alter the stored initial variables. In
contrast, from the case 2, 3 and 4 we can deduce that the templates have been
changed, either in the upper, lower or both trees.

\section{Conclusion}

We have discussed the bar-tree representation suitable for the reverse algorithm
method to extract the RoI and to label the relevant attributes in the initial
setup. The resulted patterns can be used further to automatically extract the
data from crawled targeted web pages. We argue that the method with additional
relevant web interface would reduce the administrator works significantly. On
the other hand it improves the accuracy and speed of finding the tokens and
labeling the attributes. Because the human intervention is basically required
only during the initial setup. The pattern recognition in this method is done in
an exact way without, for instance,  any predefined parameters like threshold
value etc.

We have found that this method is very effective to detect the template changes,
for instance newly inserted banners in the middle of upper or lower tree which
often occurs in any websites and leads to difficulties in existing methods. The
important point is it does not require huge computer power, nor further human
intervention once the initial setup has been done. 

Through our experimental running, we can conclude that the focused
web-harvesting deploying  the combination of reverse algorithm and bar-tree
representation is appropriate for the indirect data integration. The method
performs the data collection over targeted web sources very accurately. We also
recommend to perform the full set $\{ d_\mathrm{max}, A^\mathrm{total}, P_d, A_d
\}$ rather than the simpler set $\{ d_\mathrm{max}, A^\mathrm{total} \}$ to
obtain results with much higher accuracy in a moderate time consumption.

The real works on applying the method to further restructure the huge number of
data stored at ISI is still under progress. The  results and its effectiveness
to restructure all relevant fields in a nation wide scientific index will be
analyzed and reported elsewhere. Finally, we should remark that the method is
also applicable for the web sources in a form of list of data. The work on this
matter is also in progress.

\section*{Acknowledgment}

This work is funded by the Riset Kompetitif LIPI in fiscal year 2009 under
Contract no. 11.04/SK/KPPI/II/2009. 

\bibliographystyle{IEEEtran}
\bibliography{bartree}

\begin{IEEEbiography}[{\includegraphics[width=1in,height=1.25in,clip,
keepaspectratio]{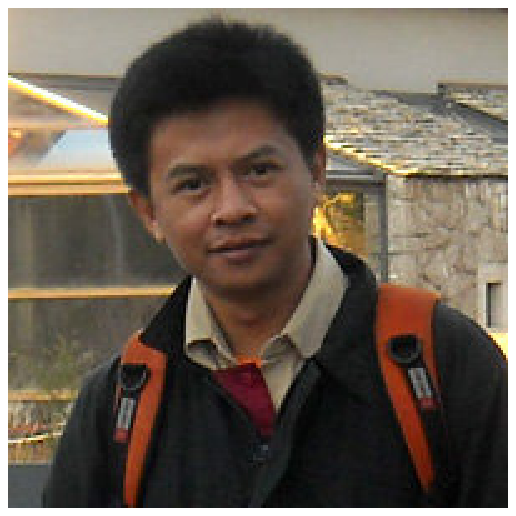}}]{ Z. Akbar}
 is a young researcher at the Group for Theoretical and Computational
Physics, Research Center for Physics, Indonesian Institute of Sciences (LIPI).
He is majoring on Computing Science, but his research is focused on the
distributing system. Since 2009 he also joined the Group for Bioinformatics and
Information Mining, Department of Computer and Information Science, University
of Konstanz as an associate researcher and working on distribution system for
bioinformatics problems. His URL is http://teori.fisika.lipi.go.id/~zaenal/.
\end{IEEEbiography}

\begin{IEEEbiography}[{\includegraphics[width=1in,height=1.25in,clip,
keepaspectratio]{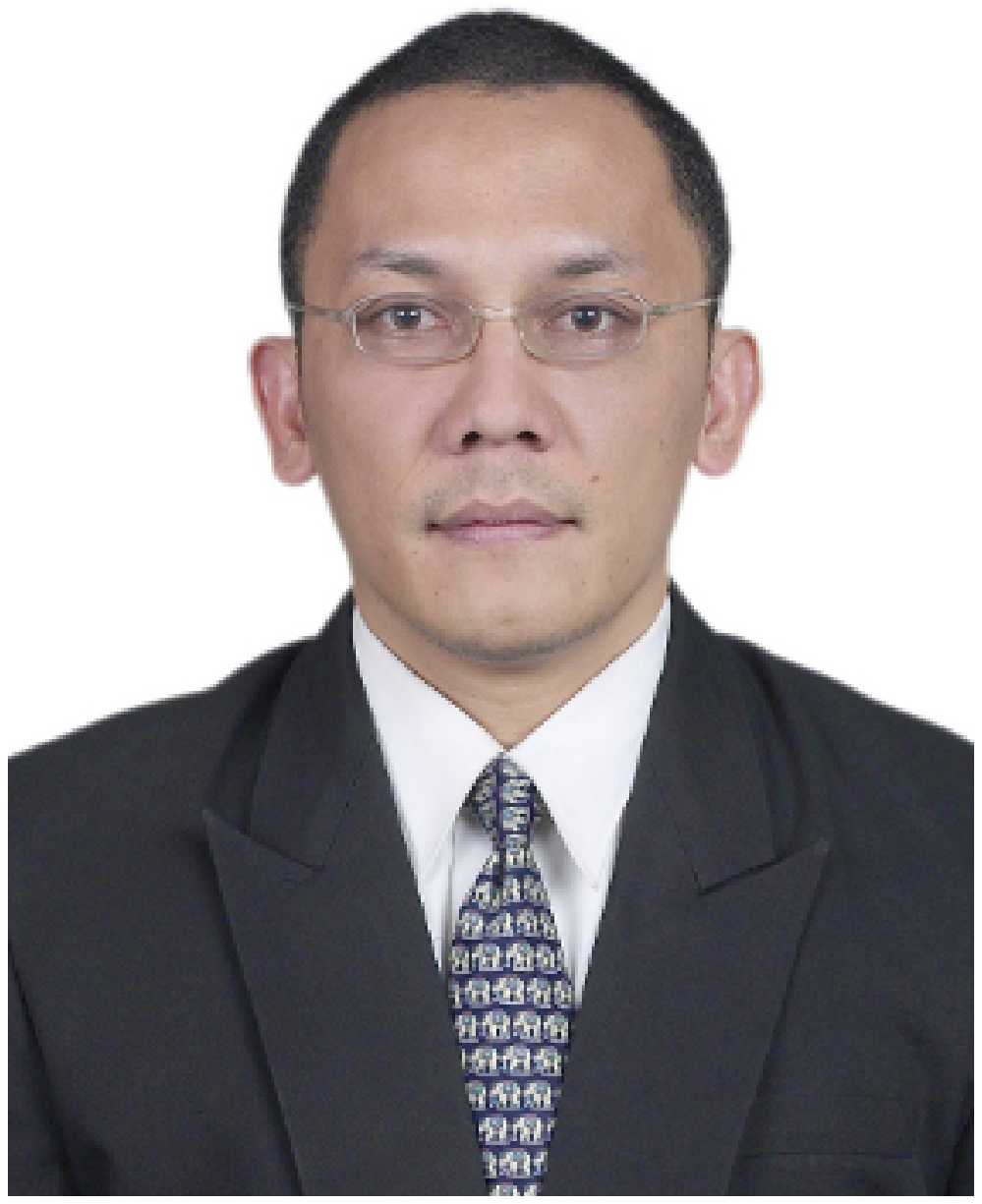}}]{L.T. Handoko}
 is a senior researcher and Group Head at the Group for Theoretical and
Computational Physics, Research Center for Physics, Indonesian Institute of
Sciences (LIPI). He is also a visiting professor at the Department of Physics,
University of Indonesia. His research interest covers broad area from
theoretical particle physics to computational science. His URL is
http://teori.fisika.lipi.go.id/~handoko/.
\end{IEEEbiography}

\end{document}